\documentclass[prx,reprint,aps,amsmath,amssymb,showpacs, superscriptaddress]{revtex4-2}
\usepackage{graphicx}

\begin{document}
\preprint{}
\draft

\title{Quantum and Classical Two-photon Interference of Single Photons with Ultralong Coherence Time}

\author{Manman Wang}\thanks{These authors contributed equally to this work}
\affiliation{Beijing Academy of Quantum Information Sciences, Beijing 100193, China.}
\affiliation{Institute of Physics, Chinese Academy of Sciences, Beijing 100190, China.}
\affiliation{University of Chinese Academy of Sciences, Beijing 101408, China.}

\author{Yanfeng Li}\thanks{These authors contributed equally to this work}
\affiliation{Beijing Academy of Quantum Information Sciences, Beijing 100193, China.}
\affiliation{Institute of Physics, Chinese Academy of Sciences, Beijing 100190, China.}
\affiliation{University of Chinese Academy of Sciences, Beijing 101408, China.}

\author{Hanqing Liu}
\affiliation{Institute of Semiconductors, Chinese Academy of Sciences, Beijing 100083, China.}
\affiliation{College of Materials Science and Opto-Electronic Technology, University of Chinese Academy of Sciences, Beijing 101408, China.}

\author{Haiqiao Ni}
\affiliation{Institute of Semiconductors, Chinese Academy of Sciences, Beijing 100083, China.}
\affiliation{College of Materials Science and Opto-Electronic Technology, University of Chinese Academy of Sciences, Beijing 101408, China.}

\author{Zhichuan Niu}
\affiliation{Institute of Semiconductors, Chinese Academy of Sciences, Beijing 100083, China.}
\affiliation{College of Materials Science and Opto-Electronic Technology, University of Chinese Academy of Sciences, Beijing 101408, China.}

\author{Xiaogang Wei}
\affiliation{Beijing Academy of Quantum Information Sciences, Beijing 100193, China.}

\author{Renfu Yang}
\affiliation{Beijing Academy of Quantum Information Sciences, Beijing 100193, China.}

\author{Chengyong Hu}\email{cyhu03@gmail.com}
\affiliation{Beijing Academy of Quantum Information Sciences, Beijing 100193, China.}

\begin{abstract}
Two-photon interference (TPI) is a fundamental phenomenon in quantum optics and
plays a crucial role in quantum information science and technology.
TPI is commonly considered as quantum interference with an upper bound
of $100\%$ for both the TPI visibility and the beat visibility in contrast to its
classical counterpart with a maximum visibility of $50\%$. However, this is not always the case.
Here we report a simultaneous observation of quantum and classical TPI of single photons with ultralong
coherence time which is longer than the photon correlation time by five orders of magnitude.
We observe a TPI visibility of $94.3\%\pm 0.2\%$ but a beat visibility of $50\%$.
Besides an anti-bunching central dip due to single-photon statistics,
we observe two bunching side peaks in cross-correlation curves for indistinguishable
photons. Using either classical wave superposition theory or quantum field approach,
we derive the same expressions for the cross-correlation functions which reproduce and
explain the experiments well. We conclude that quantum TPI with a stream of single
photons is equivalent to classical TPI, both of which are the fourth-order interference
arising from the second-order interference occurring on the time scale of photon
coherence time.

\end{abstract}

\date{\today}

\maketitle

\section{INTRODUCTION}
When two identical photons are simultaneously incident on a 50/50 beam splitter from two input ports,
they will always bind together and leave the beam splitter from the same output port. This phenomenon is called
two-photon interference (TPI) - a fundamental effect in quantum optics which was discovered in 1987 by
Hong, Ou, and Mandel (HOM) \cite{hong87}.
Since its discovery, TPI has been considered as quantum interference with no analogue
in classical physics \cite{mandel99, bouchard21}, and widely applied in quantum information science and
technology, e.g., measuring photon's bandwidth and timing\cite{hong87}, testing the degree of photon indistinguishability \cite{santori02, legero04}, Bell-state analysis/measurement \cite{bouwmeester97} and
entanglement swapping/generation \cite{pan98} for quantum
communications and networks \cite{pan12, kimble08, wehner18,lu21}, quantum information processing \cite{knill01,obrien09,pelucchi22} and
quantum metrology \cite{giovannetti11, pirandola18}.

Recently there has been great interest in the fourth-order interference between two weak
lasers (it is also called the HOM interference, or classical TPI  by some authors) for practical implementations of the measurement-device-independent quantum key distribution (MDI-QKD) \cite{lo12, ma12, xu20} which uses TPI to post-select entangled states as a time-reversed version of
entanglement-based QKD \cite{ekert91}. The HOM interference between two lasers or classical light
is usually interpreted as classical interference with a visibility less than
$50\%$ \cite{paul86, rarity05} in contrast to TPI with a maximum visibility of $100\%$ when two photons are indistinguishable \cite{hong87,mandel99,legero04}.
This visibility difference usually serves as a criterion to distinguish between quantum TPI and classical TPI.
It is well known that all photons in laser light are identical, thus
the interference visibility with an upper bound of $50\%$ underestimates
the degree of photon indistinguishability of a laser. Therefore
it is natural to ask: What's the relationship between the quantum TPI of two photons
and the HOM classical interference of two lasers? And how to characterize the genuine
photon indistinguishability of a laser?

In this work, we answer the above questions by performing
time-resolved TPI experiments in an asymmetric Mach-Zehnder interferometer (AMZI)
using single photons with ultralong coherence time.
In most TPI experiments reported so far, single photons have short coherence time
limited by the single-photon correlation time or twice the emitter's radiative lifetime.
As a result, TPI and the HOM classical interference mix up with the single-photon statistics and
it is hard to distinguish between them.
In this work we coherently
convert laser light into single photons using a single quantum dot (QD) coupling
to a doubled-sided optical microcavity in the Purcell regime. Such single photons
inherit the incident laser's ultralong coherence ($>10\ \mu$s) which is five orders
of magnitude longer than the photon correlation time, so allow us to discriminate
quantum and classical effects in TPI and related interference beat.
We observe a TPI
visibility of $94.3\%\pm 0.2\%$ but a beat visibility of $50\%$, indicating the
coexistence of quantum and classical TPI.
Besides an anti-bunching central dip due to single-photon statistics,
we observe two bunching side peaks in cross-correlation curves for indistinguishable
photons. We find these intricate phenomena are linked to each other.
Using either classical wave superposition theory or quantum field theory,
we derive the same expressions for the cross-correlation functions which reproduce and
explain the experiments well. Based on these results,
we conclude that quantum TPI with a stream of single
photons is equivalent to classical TPI, both of which are the fourth-order interference
arising from the second-order interference occurring on the time scale of photon
coherence time.

\begin{figure}[ht]
\centering
\includegraphics* [bb= 100 325 512 596, clip, width=8cm, height=5.3cm]{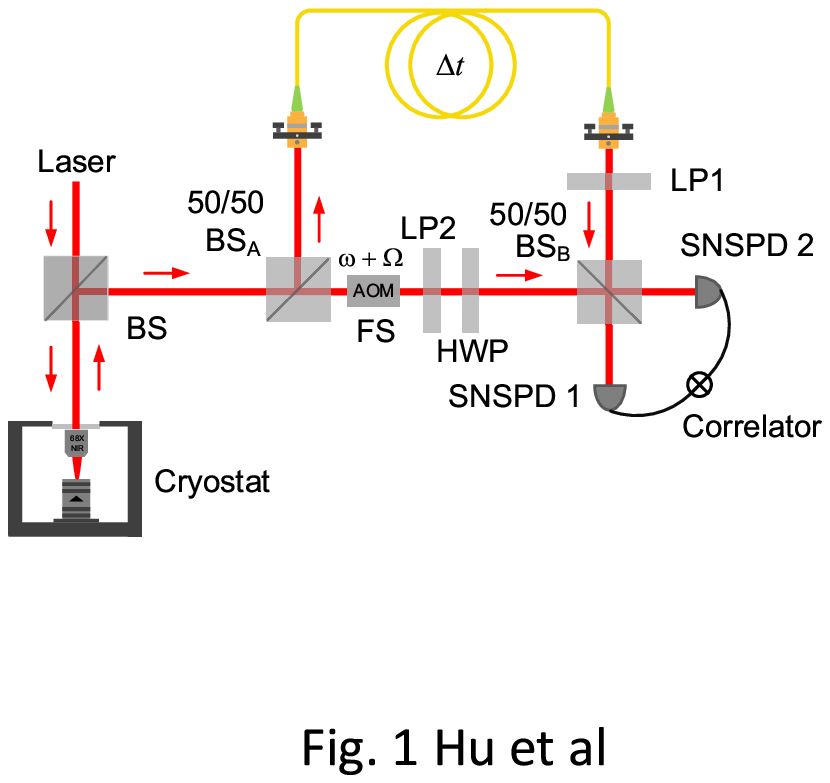}
\caption{(color online) Experimental setup for TPI measurement using a modified AMZI where
two cascaded AOM frequency shifters are placed in one arm to make tunable frequency
shifts. Either fibre or free-space
delays are used to vary the time difference $\Delta t$ between two AMZI arms.
Time-resolved correlation measurements are performed with an 8-channel time-correlated single photon counter
with a time jitter of $3\ $ps. Single photons with ultralong coherence time are generated by a single QD in an
optical microcavity placed inside a closed-cycle cryostat.
BS, $\mathrm{BS_A}$ and $\mathrm{BS_B}$: beam splitters, FS: AOM frequency shifter, LP1 and LP2: linear polarizers, HWP: half-wave plate, SNSPD1 and SNSPD2: superconducting nano-wire single-photon detectors with a time resolution of $20\ $ps.}
\label{fig1}
\end{figure}

\section{Generation of single photons with ultralong coherence time}
We designed and fabricated a pillar microcavity containing a single self-assembled In(Ga)As QD
resonantly coupling to the fundamental cavity mode with the cooperativity parameter
$C=2g^2/(\kappa \gamma_{\perp}) \gg 1$ and critical photon number
$n_0=\gamma_{\perp}\gamma_{\parallel}/(4g^2) \ll 1$, where $g$ is the QD-cavity interaction strength,
$\kappa$ is the cavity photon decay rate, $\gamma_{\parallel}$ is the QD spontaneous emission rate into leaky modes,
$\gamma_{\perp}=\gamma_{\parallel}/2+\gamma^*$ is the QD polarization decay rate, and
$\gamma^*$ is the QD pure dephasing rate.
Such design allows the incident laser light interacts with the QD deterministically,
cavity-enhanced coherent scattering and strong nonlinearity at the single-QD and single-photon
level. The cavity is defined by two mirrors made up of 18 and 30 pairs of
GaAs/Al$_{0.9}$Ga$_{0.1}$As distributed Bragg reflectors (DBRs), respectively.
The two DBR mirrors are made asymmetric in the realistic devices such that the leakage rate
from the top mirror can balance the total leakage rates from the bottom mirror,
cavity side and background absorption resided in materials.
This cavity structure mimics a double-sided cavity with zero reflectivity at the center of
the fundamental cavity mode. The details for sample growth and device fabrication can be found
in our recent work \cite{li24}.
The cavity quantum electrodynamics(CQED) parameters for the sample used in this work
are $g/2\pi=4.7\ $GHz, $\kappa/2\pi=36.8\ $GHz, $\gamma_{\parallel}/2\pi = 0.35\ $GHz and
$\gamma^*/2\pi \simeq 0\ $GHz. So the cooperativity parameter is $C=6.9$ and the critical
photon number is $n_0=6.9\times 10^{-4}$.

\begin{figure*}[ht]
\centering
\includegraphics* [bb= 130 484 448 680, clip, width=12cm, height=7.4cm]{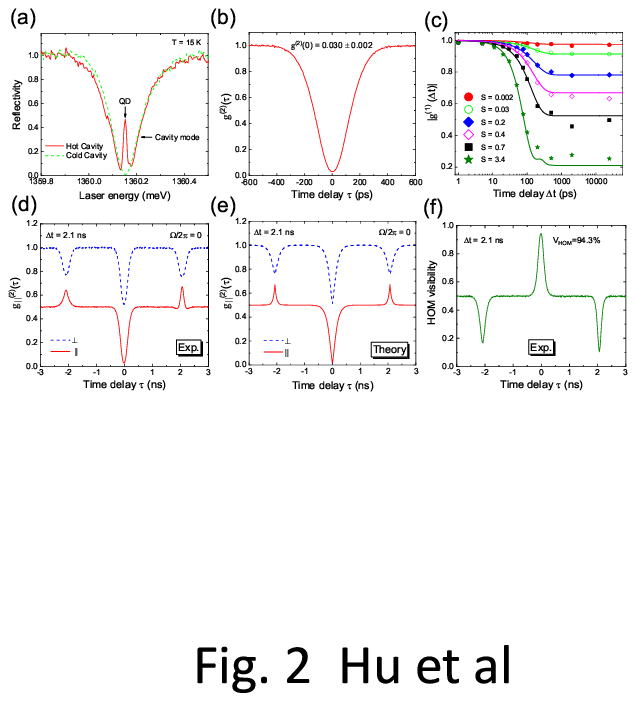}
\caption{(color online) Generation and characterization of single photons
with ultralong coherence time ($>10\ \mu$s).
(a) Reflection spectra measured by scanning the laser's wavelength. The QD-cavity coupled system (i.e., hot cavity)
behaves like a cold or empty cavity at high laser powers when the QD gets saturated by excitation.
(b) Measured second-order autocorrelation $\mathrm{g}^{(2)}(\tau)$ of
reflected light with the laser frequency fixed at the QD resonance. $\mathrm{g}^{(2)}(0)=0.030\pm 0.002$
is achieved.
(c) Measured degree of first-order coherence $|\mathrm{g}^{(1)}(\tau)|$ versus
time delay at different driving powers with a Michelson interferometer.
The solid curves are calculations using the master equation\cite{li24}.
We choose a low driving power with $S=0.01$ for TPI experiments in this work.
(d) Cross-correlation measurements for cross polarizations (blue dash)
and parallel polarizations (red solid) at $\Delta t=2.1\ $ns and $\Omega/2\pi=0\ $kHz.
The experiment time is more than half an hour.
(e) Calculated cross-correlation functions for cross polarizations (blue dash)
and parallel polarizations (red solid) at $\Delta t=2.1\ $ns and $\Omega/2\pi=0$.
(f) TPI visibility versus the time delay $\tau$ at $\Delta t=2.1$ ns and $\Omega/2\pi=0$.
The visibility peak with $V_{HOM}(0)=94.3\%\pm 0.2\%$ is accompanied by a $50\%$ background.}
\label{fig2}
\end{figure*}

The sample was placed inside a closed-cycle cryostat (see Fig. 1) and the QD transition was in resonance
with the cavity mode by temperature tuning. A cw tunable Ti:saphhire laser with the linewidth
$<100\ $kHz was used to drive the cavity. We monitor the intensity or correlation of the reflected light.
Fig. 2(a) presents the coherent reflection spectra measured by scanning the laser's wavelength.
We get nearly zero reflectivity ($R=0.89\%$) at the center frequency
of cavity mode at higher laser powers when the QD saturates. The QD transition induces a sharp reflection peak ($R=46.6\%$)
inside the cavity-mode resonance [Fig. 2(a), red solid line] at lower laser powers.

Fixing the laser frequency on the QD resonance, we measure the second-order autocorrelation
function of reflected light and achieve $\mathrm{g}^{(2)}(0)=0.030\pm 0.002$ [see Fig. 2(b)].
High single-photon purity is observed at low driving fields with the saturation parameter $S<1$.
The reflected light consists of a superposition of the driving field and the cavity output field.
The driving laser field shows Poissonian statistics while the cavity output field exhibits
super-bunching \cite{li24} due to photon-induced tunneling \cite{faraon08, majumdar12}
and multi-photon scattering \cite{shen07},
so the common picture that a single QD can only scatter (or absorb and emit) single photons
cannot be applied at low driving fields. We identify that fully destructive interference
between the driving field and the cavity output field erases the two-photon probability
amplitude in reflected light and converts the driving laser light into single photons \cite{li24},
affirming the interference picture on antibunching in resonance fluorescence proposed
40 years ago \cite{dalibard83}.

At low driving fields with $S<1$, the cavity output intensity is much weaker than the driving field, so
the first-order coherence of converted single photons is simply determined by the driving field.
Fig. 2(c) presents the degree of first-order coherence $\mathrm{g}^{(1)}(\tau)$ of reflected light versus
time delay measured with a Michelson interferometer at different driving powers. There are two coherence times observed,
$\tau_{c1} \simeq 115$ ps and  $\tau_{c2}> 24.5$ ns which is limited by the longest path delay of interferometer.
The short 115-ps coherence time which is twice the QD radiative lifetime $57\ $ps
stems from the incoherent cavity output field due to quantum fluctuations,
while the long coherence time comes from
the driving field with ultra-long coherence time ($>10\ \mu$s).
The incoherent cavity output reduces the degree of first-order coherence by its intensity fraction in reflected light,
however, this reduction becomes less significant at low driving fields.

\section{Two-photon interference in AMZI}
Next we perform TPI experiments in a modified AMZI (see Fig. 1) where two cascaded acousto-optical modulation (AOM)
frequency shifters are placed in one arm to make tunable frequency shift.
The device works at a low driving field with the saturation parameter $S=0.01$ in the Heitler regime.
The reflected single photons have a high degree of first-order
coherence with $|\mathrm{g}^{(1)}(\tau)|>95\%$ [see Fig. 2(c)] for $\tau$ less than the coherence time.
Fig. 2(d) presents the measured cross-correlation functions $\mathrm{g}^{(2)}_{\perp}(\tau)$
and $\mathrm{g}^{(2)}_{\parallel}(\tau)$ between two detectors in cross- and parallel-polarization
configurations at $\Delta t=2.1$ns and $\Omega/2\pi=0$.

In cross-polarization configuration, interference phenomena are not expected and we observe
three anti-bunching dips at $\tau=0, \pm \Delta t$ with a correlation time of $115\ $ps which is twice the
QD radiative lifetime. The central dip and two side dips arise from the single-photon correlations with
a depth of $0.5$ and $0.25$, respectively.

In parallel-polarization configuration, besides the central dip at $\tau=0$, we observe
two bunching side peaks at $\tau=\pm \Delta t$ which have not been reported before \cite{patel08, proux15}.
The HOM (or TPI) visibility is usually defined as
$V_{HOM}(\tau)=[\mathrm{g}^{(2)}_{\perp}(\tau)-\mathrm{g}^{(2)}_{\parallel}(\tau)]/\mathrm{g}^{(2)}_{\perp}(\tau)$.
From it, we get the TPI visibility $V_{HOM}(0)=94.3\%\pm 0.2\%$ [Fig. 2(f)] which is well above the $0.5$
classical limit, indicating quantum nature of TPI. The TPI visibility peak sits on a $V_{HOM}(\tau)=0.5$ background which also stems from TPI
as discussed later. For this reason, we normalize $\mathrm{g}^{(2)}_{\parallel}(\tau)$ to $0.5$ in Fig. 2(d).

The bunching side peaks persist with increasing the fibre delays up to $5\ $ km (i.e., $\Delta t<25\ \mu$s)
and then turn to antibunching dips when the fibre delays are longer than $5\ $ km (i.e., $\Delta t\geq 25\ \mu$s)
as shown in Fig. 3.

If keeping the fibre delay to $1\ $km (i.e., $\Delta t=5\ \mu$s) but varying the AOM frequency shifts,
we observe periodic changeover between bunching side peaks and antibunching side dips.
For example, the bunching side peaks
for $\Omega=0$ [see Fig. 3(b)] turn
to anti-bunching side dips inside the interference beat for $\Omega/2\pi=48.0\ $kHz [Fig. 4(a)], $\Omega/2\pi=101.6\ $kHz [Fig. 4(b)], $\Omega/2\pi=147.1\ $kHz [Fig. 4(c)], $\Omega/2\pi=246.0\ $kHz [Fig. 4(e)], but turn back to bunching side peaks for
$\Omega/2\pi=194.7\ $kHz [Fig. 4(d)].

In order to understand the above results, we adopt a general wave superposition theory (see the Supplemental Material \cite{supp}) to
calculate the cross-correlation functions for a general light field in the modified AMZI (see Fig. 1) in
cross-polarization configuration
\begin{widetext}
\begin{equation}
\mathrm{g}^{(2)}_{\perp}(\tau)=\frac{1}{N}\Biggl\{(R_A^2+T_A^2)R_BT_B\mathrm{g}^{(2)}(\tau)
+R_AT_AR_B^2\mathrm{g}^{(2)}(\tau +\Delta t)+R_AT_AT_B^2\mathrm{g}^{(2)}(\tau-\Delta t)\Biggl\},
\label{eq1}
\end{equation}
and in parallel-polarization configuration
\begin{equation}
\begin{split}
\mathrm{g}^{(2)}_{\parallel}(\tau)=& \frac{1}{N}\Biggl\{(R_A^2+T_A^2)R_BT_B\mathrm{g}^{(2)}(\tau)+R_AT_AR_B^2\mathrm{g}^{(2)}(\tau  +\Delta t)+R_AT_AT_B^2\mathrm{g}^{(2)}(\tau-\Delta t)\\
&-2R_AT_AR_BT_BV_0|\mathrm{g}^{(1)}(\tau)|^2\cos \Omega\tau\sqrt{\mathrm{g}^{(2)}(\tau-\Delta t)\mathrm{g}^{(2)}(\tau+\Delta t)}\Biggl\},
\end{split}
\label{eq2}
\end{equation}
\end{widetext}
where $N=(R_A^2+T_A^2)R_BT_B+(R_B^2+T_B^2)R_AT_A$ is the normalization factor. $R_{A.B}$ and $T_{A,B}$
are the reflection and transmission intensity coefficients of the beam splitters $\mathrm{BS_A}$ and
$\mathrm{BS_B}$ with nominal values $R_A=T_A=50\%$ and $R_B=T_B=50\%$. The parameter $V_0$ is
introduced to consider the mode overlap on $\mathrm{BS_B}$ (see the Supplemental Material \cite{supp}).
$\mathrm{g}^{(1)}(\tau)$ and $\mathrm{g}^{(2)}(\tau)$ are the first- and second-order auto-correlation functions of
the input light. $\Omega/2\pi$ is the AOM frequency shift.
Note that Eqs. (\ref{eq1}) and (\ref{eq2}) are suitable for general light
including single photons, laser light and thermal light. In case of single photons,
Eqs. (\ref{eq1}) and (\ref{eq2}) are also obtained in framework of quantum field
theory (see the Supplemental Material \cite{supp}).

A coincident event involves two photons detected by each of two detectors.
The first term in Eqs. (\ref{eq1}) and (\ref{eq2}) comes from the coincidence events for two
photons traveling through the same AMZI arm. The second and third terms count in the coincidence
events for two photons going through different arms.
The fourth term in Eq. (\ref{eq2}) is due to the fourth-order interference of two photons
which essentially originates from the second-order interference occurring on the time scale of photon
coherence time (see Supplemental Material \cite{supp}). This is the only interference term in Eq. (\ref{eq2}),
so TPI and the HOM classical interference are the same interference.
In parallel-polarization configuration, single-photon correlation contributes to the antibunching central dip with a depth of $0.5$ and
a small width of $115\ $ps, while TPI (i.e., the HOM classical interference) leads to a
broad background (i.e., the HOM dip) with a depth of $0.5$ and a width of the photon coherence time ($>10\ \mu$s).
That is why we normalize $\mathrm{g}^{(2)}_{\parallel}(\tau)$ to $0.5$ when $\Delta t \ll \tau_c$ [see Fig. 2(d) and Fig. 3].
For the same reason, the TPI visibility peak is observed on a $V_{HOM}(\tau)=0.5$ background in Fig. 2(f).
Taking $R_A=T_A=50\%$, $R_B=T_B=50\%$, $\mathrm{g}^{(2)}(0)=0.03$ and $|\mathrm{g}^{(1)}(\tau)|=\exp{(-|\tau|/\tau_c)}$ where $\tau_c$ is the photon coherence time ($>10\ \mu$s), Eqs. (\ref{eq1}) and (\ref{eq2}) well reproduce the measured $\mathrm{g}^{(2)}_{\perp}(\tau)$ and $\mathrm{g}^{(2)}_{\parallel}(\tau)$ in Fig. 2(d) with the calculated results plotted in Fig. 2(e).

To explain the bunching side peaks at $\tau=\pm \Delta t$, we rewrite Eq. (\ref{eq2}) as
\begin{widetext}
\begin{equation}
\begin{split}
\mathrm{g}^{(2)}_{\parallel}(\tau)=&\frac{1}{N}\Biggl\{(R_A^2+T_A^2)R_BT_B\mathrm{g}^{(2)}(\tau)\\
&+R_AT_AR_B\sqrt{\mathrm{g}^{(2)}(\tau+\Delta t)}\Biggl(R_B\sqrt{\mathrm{g}^{(2)}(\tau+\Delta t)}-T_BV_0|\mathrm{g}^{(1)}(\tau)|^2\cos \Omega\tau\sqrt{\mathrm{g}^{(2)}(\tau-\Delta t)}\Biggr)\\
&+R_AT_AT_B\sqrt{\mathrm{g}^{(2)}(\tau-\Delta t)}\Biggl(T_B\sqrt{\mathrm{g}^{(2)}(\tau-\Delta t)}-R_BV_0|\mathrm{g}^{(1)}(\tau)|^2\cos \Omega\tau\sqrt{\mathrm{g}^{(2)}(\tau+\Delta t)}\Biggr)\Biggl\}.
\end{split}
\label{eq3}
\end{equation}
\end{widetext}

At first we discuss the $\Omega=0$ case. If $T_B\sqrt{\mathrm{g}^{(2)}(0)}<R_BV_0|\mathrm{g}^{(1)}(\Delta t)|^2$ [refer to the third term of Eq.(\ref{eq3})], i.e.,
\begin{equation}
\Delta t<\frac{\tau_c}{2}\ln{\left[\frac{R_BV_0}{T_B\mathrm{g}^{(2)}(0)}\right]},
\label{eq4}
\end{equation}
a side peak at $\tau=\Delta t$ is expected, and otherwise we would observe a side dip at $\tau=\Delta t$.
Here we take $\mathrm{g}^{(2)}(\tau+\Delta t)=1$ at $\tau=\Delta t$ as
$\Delta t$ is much larger than the photon correlation time ($115\ $ps).

If $R_B\sqrt{\mathrm{g}^{(2)}(0)}<T_BV_0|\mathrm{g}^{(1)}(\Delta t)|^2$ [refer to the second term of Eq.(\ref{eq3})], i.e.,
\begin{equation}
\Delta t<\frac{\tau_c}{2}\ln{\left[\frac{T_BV_0}{R_B\mathrm{g}^{(2)}(0)}\right]},
\label{eq5}
\end{equation}
a side peak at $\tau=-\Delta t$ is expected, and otherwise we would see a side dip at $\tau=-\Delta t$.
Here we take $\mathrm{g}^{(2)}(\tau-\Delta t)=1$ at $\tau=-\Delta t$.

\begin{figure}[ht]
\centering
\includegraphics* [bb= 164 391 497 769, clip, width=8cm, height=9.5cm]{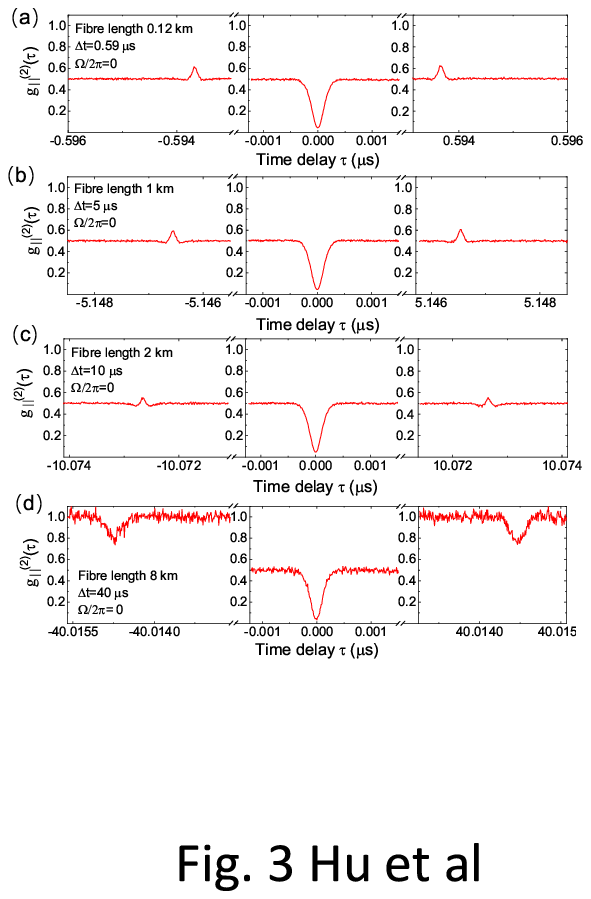}
\caption{(color online) Cross-correlation measurements in parallel-polarization configuration at different AMZI delays: (a) $0.12\ $km; (b) $1\ $km; (c) $2\ $km; (d) $8\ $km.  The AOM frequency
shift is set to zero. The driving field is kept at $S=0.01$.
The measurement time for each curve is several hours.}
\label{fig3}
\end{figure}

\begin{figure}[ht]
\centering
\includegraphics* [bb= 73 207 516 773, clip, width=7cm, height=8.9cm]{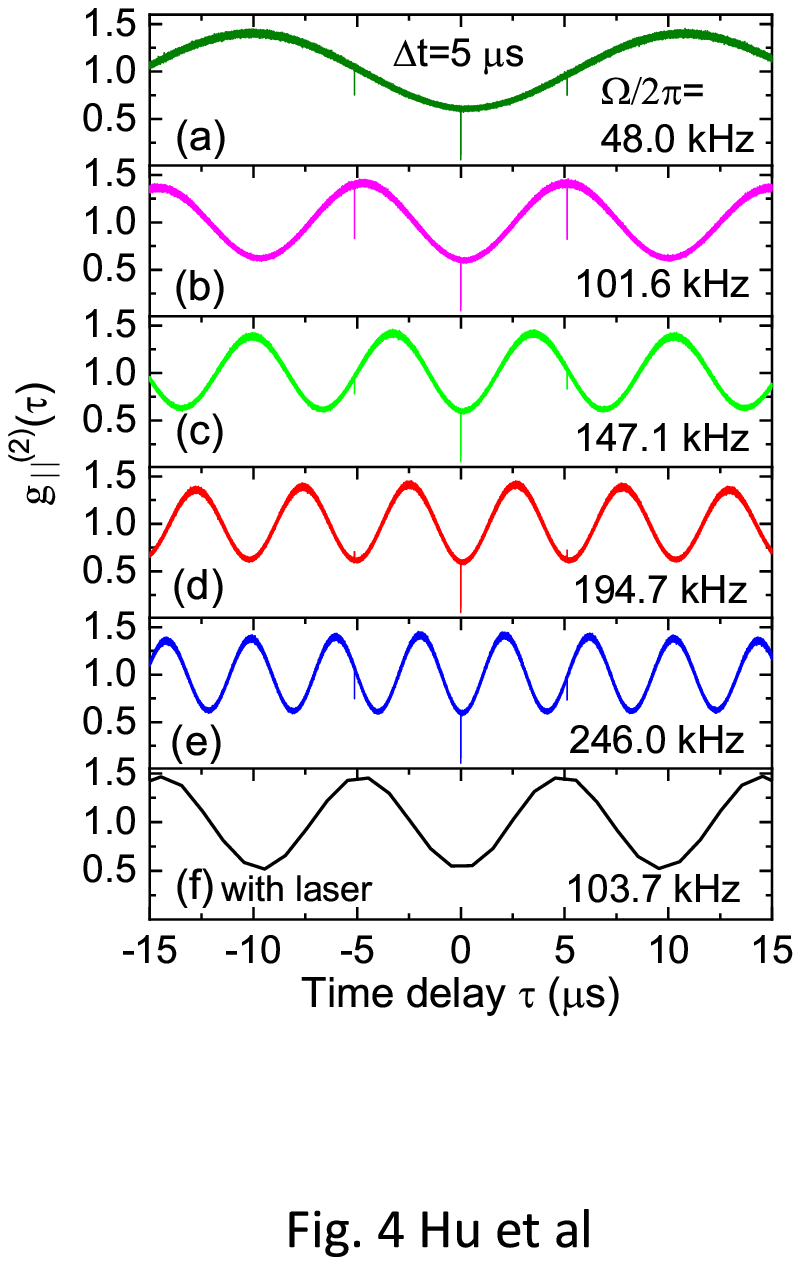}
\caption{(color online) Cross-correlation measurements with single photons in parallel-polarization configuration at different AOM frequency shifts:
(a) $\Omega/2\pi=48.0\ $kHz; (b) $\Omega/2\pi=101.6\ $kHz; (c) $\Omega/2\pi=147.1\ $kHz; (d) $\Omega/2\pi=194.7\ $kHz; (e) $\Omega/2\pi=246.0\ $kHz.
The fibre length is fixed to $1\ $km, corresponding to a time delay $\Delta t=5\ \mu$s. The driving field is kept at  $S=0.01$. Same measurement with the driving laser is plotted in (f) for comparison.}
\label{fig4}
\end{figure}

If taking $R_B=T_B=50\%$, $V_0=100\%$, and $\mathrm{g}^{(2)}(0)=0.03$, the above inequalities (\ref{eq4}) and (\ref{eq5})
reduce to $\Delta t<1.8\tau_c$. Assuming $\tau_c=10\ \mu$s, we would observe
two bunching side peaks at $\tau=\pm \Delta t$ if the AMZI fibre delays are less than $3.6\ $km (i.e., $\Delta t<18\ \mu$s),
or two anti-bunching side dips if fibre delays are longer than $3.6\ $km (i.e., $\Delta t>18\ \mu$s), in agreement with
the observations in Fig. 3 where the anti-bunching side peaks turn to dips when the fibre
length increases beyond $5\ $km. When $\Delta t \gg 1.8\tau_c$, the interference effect
cannot reach the side dips at $\tau=\pm \Delta t$ as $\mathrm{g}^{(1)}(\Delta t) \approx 0$, so
$\mathrm{g}^{(2)}_{\parallel}(\tau)$ around $\tau=\pm \Delta t$ is
normalized to one in Fig. 3(d).
The small asymmetry between the two bunching side peaks is caused by the slight difference between
$R_B$ and $T_B$, or different detection efficiencies.
In previous work \cite{proux15, patel08}, single photons generated by the QD spontaneous emission
have very short coherence time limited by twice the QD radiative lifetime. As a result,
$\Delta t>\frac{\tau_c}{2}\ln{\left[V_0/\mathrm{g}^{(2)}(0)\right]}$
is always met, so only antibunching side dips were observed in their experiments.

In the case of $\Omega\neq0$, the interference term in Eqs. (\ref{eq2}) or (\ref{eq3})
modulated by $\cos\Omega \tau$ generates an interference beat in $\mathrm{g}^{(2)}_{\parallel}(\tau)$
curves (see Fig. 4). If $\cos(\Omega\Delta t)=-1$ or $0$, we would always observe anti-bunching
side dips at $\tau=\pm\Delta t$ no matter what the fibre length is. This can be easily seen
from Eq. (\ref{eq3}).
If $\cos(\Omega\Delta t)=1$, whether side peaks or dips are observed depends on the fibre length
in the same way as the $\Omega=0$ case.
This explains the periodic changeover between side peaks and dips
at $\tau=\pm \Delta t$ when the AOM frequency shift is tuned (see Fig. 4).
Note that similar classical interference beat is also observed for the driving laser light, but
no sharp dips or peaks are observed [see Fig. 4(f)].

The beat visibility of $0.5$ in Figs. 4(a)-4(d) and the TPI visibility
background of $0.5$ in Fig. 2(f) both reflect the classical effect in TPI, while
the TPI visibility of $94.3\%\pm 0.2\%$ in Fig. 2(f) shows the quantum effect in TPI.
All these phenomena can be well explained by Eq. (\Ref{eq2}).
Taking $R_A=T_A=50\%$ and $R_B=T_B=50\%$, Eq.(\Ref{eq2}) yields $V_{HOM}(0)=V_0/[1+\mathrm{g}^{(2)}(0)]$, specially
$V_{HOM}(0)=V_0$ for a stream of single photons with $\mathrm{g}^{(2)}(0)=0$, $V_{HOM}(0)=V_0/2$ for a stream of un-correlated
photons (e.g., laser light) with $\mathrm{g}^{(2)}(0)=1$ and $V_{HOM}(0)=V_0/3$ for thermal light with $\mathrm{g}^{(2)}(0)=2$.
Therefore, $V_0=V_{HOM}(0)[1+\mathrm{g}^{(2)}(0)]$ rather than $V_{HOM}(0)$ alone
measure the true photon indistinguishability of laser light, in accord with Mandel's insight that
indistinguishability equals coherence \cite{mandel91}.

\section{DISCUSSION}
To summarize, we have demonstrated a simultaneous observation of quantum and classical
two-photon interference of single photons with ultralong coherence time which is five orders of
magnitude longer than the photon correlation time. We coherently convert laser light
into single photons using a single QD coupling to a double-sided optical microcavity in the
Purcell regime. We observe a TPI visibility of $94.3\%\pm 0.2\%$ but a beat visibility
of $50\%$. Besides the anti-bunching central dip due to single-photon statistics,
we also observe two bunching side peaks in cross-correlation curves for indistinguishable
photons due to second-order interference and intricate photon correlations.
We reproduce the experimental results using either classical wave superposition theory
or quantum field approach. We conclude that quantum TPI with a stream of single photons
is equivalent to classical TPI, both of which are the
fourth-order interference arising from the second-order interference occurring
on the time scale of photon coherence time. Our work sheds new light on the nature of TPI.

Moreover, we point out that the technique to convert laser light into single photons can
be utilized to measure the genuine photon indistinguishability of lasers, which is not
accessible before.
Inheriting the laser's first-order coherence time and robust photon indistinguishability,
laser-converted single photons \cite{li24} could be a key resource for interference-based
quantum information technologies, e.g., to establish
high-quality TPI between remote quantum memories \cite{zhai22} or to realize long-distance TPI \cite{you22},
an important step towards quantum internet \cite{kimble08, wehner18, lu21}.

\begin{acknowledgments}
This work is supported by the Beijing Natural Science Foundation
under the grant IS23069. Z.C. Niu is grateful to the National Key Technology R$\&$D Program of China
under the grant 2018YFA0306101 for financial support.
\end{acknowledgments}

\end{document}